\documentclass[aps,prl,showpacs,twocolumn,letterpaper]{revtex4}
\usepackage{graphicx}
\usepackage{amsmath}
\def\br{\underline{\bf r}}
\def\bR{\underline{\bf R}}

\def\bx{\underline{\bf x}}

\def\ext{_{\rm ext}}

\def\half{\frac{1}{2}}
\def\quart{\frac{1}{4}}

\def\la{{\langle\, }}
\def\ra{{\,\rangle }}

\def\hT{{\hat T}}
\def\hS{{\hat S}}
\def\hV{{\hat V}}
\def\hU{{\hat U}}
\def\hH{{\hat H}}
\def\hW{{\hat W}}
\def\hp{{\hat p}}

\def\hr{{\hat{{\bf r}}}}
\def\hP{{\hat P}}
\def\hR{{\hat{{\bf R}}}}
\def\hv{{\hat v}}
\def\hj{{\hat j}}
\def\hJ{{\hat J}}
\def\hbr{{\hat{\bf r}}}

\def\hbJ{{\hat{\bf J}}}

\def\hn{{\hat n}}
\def\hN{{\hat N}}

\def\bxt{{({\bf x}, t)}}
\def\brt{{({\bf r}, t)}}
\def\bRt{{({\bf R}, t)}}

\def\ubxt{{(\underline{{\bf x}}, t)}}
\def\ubrt{{(\underline{{\bf r}}, t)}}
\def\ubRt{{(\underline{{\bf R}}, t)}}

\def\bA{{\bf A}}

\def\sn{_{\rm n}} % sub n
\def\se{_{\rm e}} % sub e
\def\n{{\rm n}}
\def\ee{_{\rm ee}}
\def\en{_{\rm en}}
\def\nuce{_{\rm ne}}
\def\nn{_{\rm nn}}
\def\exte{_{\rm ext,e}}
\def\extn{_{\rm ext,n}}
\def\Ne{{N_{\rm e}}}
\def\Nn{{N_{\rm n}}}

\def\rel{{\bf \underline{r} }}
\def\rnucl{{\bf \underline{R} }}

\def\sph_int{ {\int d^3 r}}

\begin{document}

\title{Stochastic Quantum Molecular Dynamics}
\author{Heiko Appel}
\email[Electronic address:\;]{appel@physics.ucsd.edu}
\author{Massimiliano Di Ventra}
\email[Electronic address:\;]{diventra@physics.ucsd.edu}
\address{University of California San Diego, La Jolla,
         California 92093, USA}
\date{\today}

%%%%%%%%%%%%%%%%%%%%%%%%%%%%%%%%%%%%%%%%%%%%%%%%%%%%%%%%%%%%%%%%%%
%%                          Abstract                             %
%%%%%%%%%%%%%%%%%%%%%%%%%%%%%%%%%%%%%%%%%%%%%%%%%%%%%%%%%%%%%%%%%%
\begin{abstract}
An approach to correlated dynamics of quantum nuclei and electrons
both in dynamical interaction with external environments is
presented. This stochastic quantum molecular dynamics rests on a
theorem that establishes a one-to-one correspondence between the
total ensemble-averaged current density of interacting nuclei and
electrons and a given external vector potential. The theory allows
for a first-principles description of phenomena previously
inaccessible via standard quantum molecular dynamics such as
electronic and nuclear relaxation in photochemistry, dissipative
correlated electron-ion dynamics in intense laser fields, nuclear
dephasing, etc. As a demonstration of the approach, we discuss the
rotational relaxation of 4-(N,N-dimethylamino)benzonitrile in a
uniform bath in the limit of classical nuclei.
\end{abstract}

% 71.15.-m Methods of electronic structure calculations
% 31.70.Hq Time-dependent phenomena: excitation and relaxation
% 31.15.ee Time-dependent density functional theory
% 03.65.Yz Decoherence; open systems; quantum statistical methods

\maketitle

%%%%%%%%%%%%%%%%%%%%%%%%%%%%%%%%%%%%%%%%%%%%%%%%%%%%%%%%%%%%%%%%%%
%%                          INTRODUCTION                         %
%%%%%%%%%%%%%%%%%%%%%%%%%%%%%%%%%%%%%%%%%%%%%%%%%%%%%%%%%%%%%%%%%%
Quantum molecular dynamics (QMD) approaches, as formulated, e.g., in
the Born-Oppenheimer, Ehrenfest or Car-Parrinello methods, have
proved to be extremely useful tools to study the dynamics of
condensed systems \cite{MH00, DM02, MU06}. In these approaches, the
time-evolution of the nuclear degrees of freedom is described by
classical Newton-type equations, and the forces acting on the nuclei
are typically derived from the electronic wave-function at the
instantaneous nuclear configuration. By construction,
Born-Oppenheimer and Car-Parinello QMD refer to the electronic
ground-state energy, whereas Ehrenfest QMD also allows to take
excited electronic wave-functions into account.

In these approaches, energy dissipation and thermal coupling to the
environment are usually described with additional thermostats
coupled directly to the classical nuclear degrees of freedom.
Alternatively, Langevin dynamics can be employed for the
time-evolution of the classical nuclei \cite{FS01}. Other routes
that take environment effects into account are provided by the
so-called quantum mechanics/molecular mechanics (QM/MM) methods or
by implicit solvation or continuum models (see, e.g., \cite{ST07}).
Here, the system of interest is treated quantum mechanically (e.g.,
with one of the above QMD methods) and the environment is treated
explicitly within classical molecular mechanics or, implicitly, via
an electrostatic continuum model.

However, all of the above approaches share the common feature that
the electronic subsystem is treated always as a {\em closed} one,
namely it cannot exchange energy and momentum with the
environment(s). This is a major limitation since in many physical
situations, electrons, being lighter particles than nuclei, respond
much faster than the latter ones to dynamical changes in their
surrounding. This can lead to electronic transitions among excited
states (manifested, e.g., in an effective electron temperature
rather different than the ionic one~\cite{Rob2008}), which in turn
may result in different ionic forces compared to the forces obtained
from a closed electronic quantum system at (typically) zero
temperature. An even more complicated situation may arise (for
instance, for light nuclei in intense laser fields) when the quantum
nature of nuclei correlates with the electron dynamics, with the
external environment(s) mediating transitions between electron-ion
correlated states.

In this paper we propose a novel approach that allows to treat both
the electronic and (in principle, quantum) nuclear degrees of
freedom, open to one or more environments. Our approach, which we
term stochastic quantum molecular dynamics (SQMD), is based on a
stochastic current density-functional theory for the combined
dynamics of quantum electrons and nuclei. The theorem of stochastic
time-dependent current-density-functional theory for a single
particle species has been proved in Ref.~\cite{DiVDA07,DiVDA08}.
Here we extend it to its non-trivial multi-species case. In the
limit of classical nuclei, this formulation reduces to a molecular
dynamics scheme which couples the nuclear degrees of freedom to an
open electronic system that can exchange energy and momentum with
the environment.

%%%%%%%%%%%%%%%%%%%%%%%%%%%%%%%%%%%%%%%%%%%%%%%%%%%%%%%%%%%%%%%%%%
In the following, we consider a system of $\Ne$ electrons with
coordinates $\br \equiv \{ {\bf r}_j \}$ and $N_n = \sum_s N_{s,n}$
nuclei, where each species $s$ comprises $N_{s,n}$ particles with
charges $Z_{s,j}$, masses $M_{s,j}$, $j=1\ldots N_{s,n}$, and
coordinates $\bR \equiv \{ {\bf R}_{s,j} \}$, respectively. To
simplify the notation we denote with $\bx \equiv \{ \bR, \br \}$ the
combined set of electronic and nuclear coordinates and we use the
combined index $\alpha = \{s,j\}$ for the nuclear species. Now,
suppose that such a system, subject to a classical electromagnetic
field, whose vector potential is $\bA(t)$, is described by the
many-body Hamiltonian $\hat H_S(t)$
\begin{equation}
\begin{split}
\label{eq:TotalHamiltonian} \hH_S(t) =
  & \hT\se(\rel, t)   + \hW\ee(\rel)
   + \hU\exte(\rel, t) + \hW\en(\rel,\rnucl)   \\
  & + \hT\sn(\rnucl, t) + \hW\nn(\rnucl) + \hU\extn(\rnucl, t),
\end{split}
\end{equation}
where $\hT\se(t)$ and $\hT\sn(t)$ are the kinetic energies of
electrons and ions, with velocities $\hv_k(t) = [ \hp_k +
e\bA(\hr_k, t) ]/m$ and $\hV_\alpha(t) = [ \hP_\alpha - Z_\alpha
\bA(\hR_\alpha, t) ]/M_\alpha$, respectively. The electron-electron,
electron-nuclear and nuclear-nuclear interaction terms take the form
\begin{equation}
\begin{split}
\hW\ee(\rel) &= \frac{1}{4\pi\epsilon_0}\sum_{{j < k}}^\Ne
  \frac{e^2}{|\hr_j - \hr_k|}
\equiv  \sum_{{j < k}}^\Ne w\ee(\hr_j - \hr_k),\\
\hW\nn(\rnucl) &= \frac{1}{4\pi\epsilon_0}\sum_{ \alpha < \beta }^\Nn
  \frac{Z_\alpha\,Z_\beta}{|\hR_\alpha - \hR_\beta|}
\equiv  \sum_{{\alpha < \beta}}^\Nn w\nn(\hR_\alpha - \hR_\beta),\\
\hW\en(\rel,\rnucl) &=
 - \frac{1}{4\pi\epsilon_0}\sum_{k=1}^\Ne
   \sum_{\alpha=1}^\Nn \frac{e\,Z_\alpha}{|\hr_k - \hR_\alpha|} \\
 & \equiv \sum_{k=1}^\Ne \sum_{\alpha=1}^\Nn w\en(\hr_k -
 \hR_\alpha),
\end{split}
\end{equation}
and we also allow for external time-dependent scalar potentials
$\hU\exte(\rel, t)$ and $\hU\extn(\rnucl, t)$ that act on electronic
and nuclear degrees of freedom, respectively. We then let our system
interact with one or more environments represented by a dense
spectrum of bosonic degrees of freedom, described by the Hamiltonian
$\hat H_B(t)$. The system and the environments are coupled by a
bilinear interaction term $\hat H_{SB} = \sum_j \hat S_j \hat B_j$.
Here $\hat S$, $\hat B$, refer to subsystem and bath operators,
respectively~\footnote{Note that both the electron-phonon and the
electron-photon interaction can be cast in the form of $\hat
H_{SB}$.}. The total Hamiltonian $\hat H(t)$ of system and
environment then takes the form
\begin{equation}
\hat H(t) = \hat H_S(t) + \hat H_B(t) + \sum_j \hat S_j \hat B_j.
\label{eq:Htotal}
\end{equation}
%
%%%%%%%%%%%%%%%%%%%%%%%%%%%%%%%%%%%%%%%%%%%%%%%%%%%%%%%%%%%%%%%%%%
By projecting out the bath degrees of freedom, one then obtains the
reduced dynamics of the system of interest~\cite{GN99}. While, quite
generally, we could work with the resulting non-Markovian dynamics,
for the purpose of this paper, we further make a memory-less
approximation for the bath. This leads us to consider the following
stochastic many-body Schr\"odinger equation
($\hbar=1$)~\footnote{The reasons why we work with a stochastic
Schr\"odinger equation and not with a density-matrix approach are
due to both the possible loss of positivity of the density matrix
during dynamics, and the fact that a Kohn-Sham Hamiltonian depends
on the density and/or current density, and is thus generally
stochastic~\cite{DiVDA07,DiVDA08}.}
\begin{equation}
\label{eq:SSE} i\partial_t \Psi(\bx, t) =
   {\hat H_S}(t)\Psi(\bx, t)
    - \half i {\hat S^\dagger}{\hat S}\Psi(\bx, t)
   + l(t){\hat S} \Psi(\bx, t),
\end{equation}
where, for simplicity, from now on we consider only the coupling
$\hat S$ to a single bath, which could be position and/or time
dependent. The function $l(t)$ describes a stochastic process with
zero ensemble average $\overline{l(t)} = 0$ and
$\delta$-autocorrelation $\overline{l(t)l(t')} = \delta(t-t')$. Here
$\overline{\cdots}$ describes the statistical average over all
members of an ensemble of identical systems with a common initial
state $\Psi(\bx, t=0)$ (which need not be pure) evolving under
Eq.~(\ref{eq:SSE}). Given the set of potentials $\hU\ext(\bx, t) =
\hU\exte(\rel, t) + \hU\extn(\rnucl, t)$ and $\bA(t)$ one can always
find a gauge transformation $\Lambda(\bx, t)$ so that the scalar
potentials vanish at all times, and implying that $\bA(t) \equiv
\bA(\bx, t)$. In the following we assume that such a gauge
transformation has been performed.

%%%%%%%%%%%%%%%%%%%%%%%%%%%%%%%%%%%%%%%%%%%%%%%%%%%%%%%%%%%%%%%%%%
At this stage we face a large number of choices for the definition
of densities for the construction of a density-functional
description of this problem. In fact, certain groupings of nuclear
particle and current densities might be better suited to specific
physical situations. To maintain flexibility we do not specialize at
this point and we prove a theorem for the total current density of
electrons and nuclei. To that end, we introduce electronic and
nuclear current operators in terms of the velocity fields $\hv_k(t)$
and $\hV_\alpha(t)$
\begin{equation}
\hj\brt = \frac{e}{2m} \sum_k \{\hv_k(t),\delta(\bf{r} - \hbr_k) \}
\end{equation}
\begin{equation}
\hJ_\alpha\bRt =
   \frac{Z_\alpha}{2M_\alpha}\hspace{-5mm}\sum_{\stackrel{\beta}
       {Z_\alpha = Z_\beta,\, M_\alpha = M_\beta}}\hspace{-5mm}
\{\hV_\beta(t), \delta(\bf{R} - {\hat{\bf R}}_\beta) \}
\end{equation}
where $\{\hat p, \hat q\} = \hat p \hat q + \hat q \hat p$ denotes
the usual anticommutator. Likewise, the usual charge density
operators can be defined as $\hn\brt = \sum_k \delta(\bf{r} -
\hbr_k)$ for the electrons and $\hN_\alpha\bRt =
  \sum_{\beta; Z_\alpha = Z_\beta,\, M_\alpha = M_\beta}
 \delta(\bf{R} - {\hat{\bf R}}_\beta)
$ for each nuclear species. The total particle and current density
operators of the system can then be written as $\hN\bxt = \hn\brt +
\sum_\alpha \hN_\alpha\bRt$ and $ \hbJ\bxt = \hj\brt + \sum_\alpha
\hJ_\alpha\bRt$, respectively.

Solutions of the stochastic Schr\"odinger equation,
Eq.~(\ref{eq:SSE}), leads to an ensemble of stochastic quantum
trajectories which have ensemble-averaged total charge $
\overline{N(\bx, t)} =
   \overline{\la \hN(\bx, t) \ra}$ and current
densities $\overline{{\bf J}(\bx, t)} =
   \overline{\la \hbJ(\bx, t) \ra}$.
Here $\la\cdots\ra$ denotes the quantum mechanical average.

%%%%%%%%%%%%%%%%%%%%%%%%%%%%%%%%%%%%%%%%%%%%%%%%%%%%%%%%%%%%%%%%%%
Contrary to the multi-component formulation of Kreibich {\em et al.}
\cite{KG01,KLG08} we will not employ a body-fixed frame for the
solution of Eq.~(\ref{eq:SSE}). This is motivated by the fact that
for the initial-value problem of Eq.~(\ref{eq:SSE}) a general
initial condition for the state vector and, in addition, a general
time- and space-dependent external vector potential $\bA(\bx,t)$
would break the translational and rotational invariance of the
original operator $\hat H_S(t) - \hU\ext(\bx, t)$ in
Eq.~(\ref{eq:TotalHamiltonian}). For our purposes it will therefore
be sufficient to work in the laboratory frame. We have now collected
all ingredients to state the following result.

%%%%%%%%%%%%%%%%%%%%%%%%%%%%%%%%%%%%%%%%%%%%%%%%%%%%%%%%%%%%%%%%%%
%%%%%%%%%%%                  Theorem                    %%%%%%%%%%
%%%%%%%%%%%%%%%%%%%%%%%%%%%%%%%%%%%%%%%%%%%%%%%%%%%%%%%%%%%%%%%%%%
{\em Theorem.---} For a given bath operator $\hat S$, many-body
initial state $\Psi(\bx, t=0)$ and external vector potential $
\bA(\bx,t)$, the dynamics of the stochastic Schr\"odinger equation
in Eq.~(\ref{eq:SSE}) generates ensemble-averaged total particle and
current densities $\overline{N(\bx, t)}$ and $\overline{{\bf J}(\bx,
t)}$. Under reasonable physical assumptions, any other vector
potential $\bA'(\bx,t)$ (but same initial state and bath operator)
that leads to the same ensemble-averaged total particle and current
density, has to coincide, up to a gauge transformation, with
$\bA(\bx,t)$~\footnote{As in~\cite{DiVDA07,DiVDA08} we are
implicitly assuming that given an initial condition, bath operator,
and ensemble-averaged current density, a unique ensemble-averaged
density can be obtained from its equation of motion. Therefore, the
density is not independent of the current density, and our theorem
establishes a one-to-one correspondence between current density and
vector potential.}.

%%%%%%%%%%%%%%%%%%%%%%%%%%%%%%%%%%%%%%%%%%%%%%%%%%%%%%%%%%%%%%%%%%
%%%%%%%%%%%                  Proof                      %%%%%%%%%%
%%%%%%%%%%%%%%%%%%%%%%%%%%%%%%%%%%%%%%%%%%%%%%%%%%%%%%%%%%%%%%%%%%
{\em Proof.---} The strategy for the proof of this statement follows
the technique introduced in Ref.~\cite{V04} and also employed in
Ref.~\cite{DiVDA07}. It is thus sufficient to provide the relevant
equation of motion and summarize the rationale behind the procedure.
The relevant equation of motion is the one for the ensemble-averaged
current density
\begin{equation}
\begin{split}
\label{eq:elCurrentEOM} &m \frac{\partial}{\partial t}
\overline{j_\alpha\ubrt} + \sum_\alpha M_\alpha
\frac{\partial}{\partial t} \overline{J_\alpha\ubRt} =
   \overline{ N\ubxt } \frac{\partial}{\partial t} \bA\ubxt \\[-2mm]
& - \overline{{\bf J}\ubxt} \times (\nabla \times \bA\ubxt)
  + \overline{\la {\cal{\hat F\ee}}\ubxt \ra }
  + \overline{\la {\cal{\hat F\en}}\ubxt \ra } \\
& + \overline{\la {\cal{\hat F\nn}}\ubxt \ra }
  + m \overline{ \la {\cal{\hat G}}\ubrt \ra }
  + \sum_\alpha {M_\alpha}
      \overline{ \la {\cal{\hat G_{\n,\alpha}}}\ubRt \ra }.
\end{split}
\end{equation}
Here we have introduced the electron-electron, electron-nuclear
and nuclear-nuclear interaction densities
\begin{eqnarray}
{\cal{\hat F\ee}}\ubxt &=& - \sum_{i\ne j} \delta({\bf r} -{\bf r}_j)
        \nabla_j w\ee(r_i - r_j) + m\nabla\cdot\hat\sigma\ee \nonumber \\
{\cal{\hat F\en}}\ubxt &=& - \sum_{i\ne j} \delta({\bf x} -{\bf x}_j)
        \nabla_j w\nn(x_i - x_j) + m\nabla\cdot\hat\sigma\en  \nonumber \\
&& + \sum_\alpha M_\alpha\nabla\cdot\hat\sigma\nuce,
\end{eqnarray}
where ${\cal{\hat F\nn}}$ can be obtained from ${\cal{\hat F\ee}}$
with the replacements ${\bf r} \rightarrow {\bf R}$, $w\ee
\rightarrow w\nn$, $m\nabla\cdot\hat\sigma\ee  \rightarrow
\sum_\alpha M_\alpha \nabla_\alpha\cdot\hat\sigma\nn$. The stress
tensors are
\begin{equation}
\begin{split}
\hat \sigma{\ee}_{\,i,j} =&
   - \quart \sum_k \{ v_i, \{v_j,\delta({\bf r} - r_k)\} \} \\
\hat \sigma{\en}_{\,i,j} =&
   - \quart \sum_k \{ v_i, \{V_j,\delta({\bf R} - R_k)\} \}, \\
\end{split}
\end{equation}
with $\hat \sigma{\nuce}_{\,i,j}$ and $\hat \sigma{\nn}_{\,i,j}$
obtained by the exchange ${\bf v} \leftrightarrow {\bf V}$. The
remaining terms in Eq.~(\ref{eq:elCurrentEOM}) describe the
electromagnetic contribution (terms containing $\bA$) and the force
densities due to the bath
\begin{equation}
\begin{split}
{\cal{\hat G}}\ubrt = \hS^\dagger\hj\ubrt\hS
 - \half \{ \hS^\dagger\hS, \hj\ubrt \} \\
{\cal{\hat G_\alpha}}\ubRt = \hS^\dagger\hJ_\alpha\ubRt\hS
 - \half \{ \hS^\dagger\hS, \hJ\ubRt \}. \\
\end{split}
\end{equation}

%%%%%%%%%%%%%%%%%%%%%%%%%%%%%%%%%%%%%%%%%%%%%%%%%%%%%%%%%%%%%%%%%%
We can now consider another (primed) system with different initial
condition, interaction potentials $w'\ee$, $w'\nn$, $w'\en$ and
external vector potential $\hat \bA'(\bx,t)$. By taking the
difference of the equations of motion for the total current
densities $\overline{{\bf J}(\bx, t)}$ and $\overline{{\bf J'}(\bx, t)}$ in the
unprimed and primed system we arrive at an equation of motion for
$\Delta \bA (\bx,t) = \bA(\bx,t)-\bA'(\bx,t)$. The difference of the
vector potentials $\Delta \bA (\bx,t)$ can then be shown to be
completely determined by the initial condition and a series
expansion in time about $t=0$. If the two total current densities of
the primed and unprimed system coincide then the unique solution -
up to a gauge transformation - is given by $\Delta \bA (\bx,t) = 0$,
when the initial conditions and interaction potentials are the same
in the two systems, which proves the theorem.

%%%%%%%%%%%%%%%%%%%%%%%%%%%%%%%%%%%%%%%%%%%%%%%%%%%%%%%%%%%%%%%%%%
%%%%%%%%%%%                Discussion                   %%%%%%%%%%
%%%%%%%%%%%%%%%%%%%%%%%%%%%%%%%%%%%%%%%%%%%%%%%%%%%%%%%%%%%%%%%%%%
{\em Discussion.---} With this theorem we can now set up a Kohn-Sham
(KS) scheme of SQMD where an exchange-correlation (xc) vector
potential ${\bf A}_{xc}$ (functional of the initial state, bath
operator, and ensemble-averaged current density) acting on
non-interacting species, allows to reproduce the exact
ensemble-averaged density and current densities of the original
interacting many-body system. The resulting charge and current
densities would thus contain all possible correlations in the system
- if we knew the exact functional. As mentioned before, alternative
schemes could be constructed (based on corresponding theorems that
could be proved as the above one) by defining different densities
and current densities (e.g., by lumping all nuclear densities into
one quantity~\cite{KG01}). While this may seem a drawback, it is in
fact an advantage: some of those schemes may be more appropriate for
specific physical problems. Irrespective, one would need to
construct xc functionals for the chosen scheme. While this program
is possible, it is beyond the scope of the present paper. Therefore,
as a first practical implementation of the proposed SQMD, we
consider the limit of classical nuclei. This is definitely the
simplest case, but it is by no means trivial, and by itself already
shows the wealth of physical problems one can tackle with SQMD and
the physics one can can extract from it.

\begin{figure}
  {\par\centering
  \vspace{-5mm}
  \resizebox*{0.5\columnwidth}{!}{
     \includegraphics{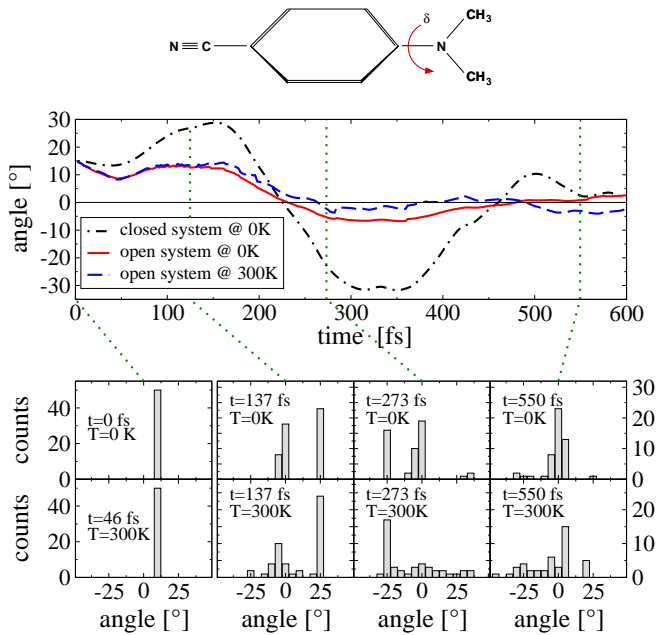}
   }\\[3mm]
  \resizebox*{\columnwidth}{!}{
     \includegraphics{dihedral-average+bins-4x2}}
  \par}
  \caption{(Color online)
  Upper panel: schematic illustration of
               4-(N,N-dimethylamino)benzonitrile and the
               dihedral angle $\delta$.
  Center panel: comparison between the dihedral angle $\delta$
               as a function of time for a closed quantum system
               (dashed line) at 0 K and an open quantum system
               (solid line) at 300 K.
  Lower panel: Distributions of the
               dihedral angle $\delta$ for 50 members of
               the stochastic ensemble for bath temperatures of
               0 K and 300 K at different times (as
               indicated by the dotted connection lines between the
               center and the lower panel).
  \label{fig:dihedral_angle}}
\end{figure}

In the limit of classical nuclei we then solve the stochastic
time-dependent KS equations with given KS Hamiltonian
$H_{KS}(\bx,t)$ for the electronic degrees of freedom, at each
instantaneous set of nuclear coordinates. The result provides the KS
Slater determinant $\Psi_{KS}(\bx,t)$ from which the forces on
nuclei are computed as ${\bf F}_{\alpha}=-\langle
\Psi_{KS}(\bx,t)|\nabla_{\alpha} \hat H_{KS}(\bx,t)| \Psi
_{KS}(\bx,t)\rangle$, for each realization of the stochastic
process. One then has direct access to both the average dynamics as
well as its {\em distribution}~\footnote{We have implemented a
quantum-jump algorithm \cite{BP02} in the real-space real-time
package {\tt octopus} \cite{CMA06} to integrate the KS equations of
motion. We have used the bath operators as in Ref.~\cite{PDDiV08},
and the adiabatic local-density approximation for the xc potential.
We note that the jumps introduced by the bath in the quantum-jump
algorithm appear like a ``surface hopping''. The relaxation rates
derived from $\hat H_{SB}$ and the jumps induced by the bath,
however, provide a solid framework for a surface hopping scheme.
More details of the numerics will be given in a forthcoming
publication.}. As application we consider the rotational relaxation
of 4-(N,N-dimethylamino)benzonitrile (DMABN) in a uniform bath. The
relaxation rate that enters the definition of the operator $\hat S$
can be derived in principle from the system-bath interaction term
$\hat H_{SB}$ in Eq.~(\ref{eq:Htotal})~\cite{GN99}. In the case of
DMABN, relaxation rates are available from experiment \cite{DE06},
so that we choose $\tau=100$\,fs for our simulation which is within
the experimentally observed magnitude. We choose as initial state
for the SQMD simulation a rotated dimethyl side-group ($\delta =
15^{\circ}$) of DMABN. In Fig.~\ref{fig:dihedral_angle} we show the
average angle and the binned angle distribution of the dihedral
angle $\delta$ for 50 stochastic realizations for bath temperatures
of 0K and 300K as a function of time and compare with the closed
system solution. The angle distributions in the lower panel of
Fig.~\ref{fig:dihedral_angle} clearly show the temperature-dependent
relaxation of the rotational motion in the open system case. Most
importantly, from the transient dynamics of the angle distributions,
it is also apparent that the system is not relaxing uniformly to the
equilibrium configuration $\delta = 0$. Rather, it approaches
equilibrium via a series of quasi-bimodal distributions, with the
higher temperature ``smoothing'' these distributions. We also
emphasize that the damping of the nuclear motion originates in our
simulation exclusively from the forces that are calculated from the
electronic open quantum system wave-functions. No additional
friction term has been added to the nuclear equation of motion. It
would be thus interesting to verify such a prediction with available
experimental capabilities.

%%%%%%%%%%%%%%%%%%%%%%%%%%%%%%%%%%%%%%%%%%%%%%%%%%%%%%%%%%%%%%%%%%
%%%%%%%%%%%                  Figure 1                   %%%%%%%%%%
%%%%%%%%%%%%%%%%%%%%%%%%%%%%%%%%%%%%%%%%%%%%%%%%%%%%%%%%%%%%%%%%%%
%%%%%%%%%%%%%%%%%%%%%%%%%%%%%%%%%%%%%%%%%%%%%%%%%%%%%%%%%%%%%%%%%%
%%%%%%%%%%%                  Summary                    %%%%%%%%%%
%%%%%%%%%%%%%%%%%%%%%%%%%%%%%%%%%%%%%%%%%%%%%%%%%%%%%%%%%%%%%%%%%%
In summary, we have presented a novel quantum molecular dynamics
approach which we term stochastic quantum molecular dynamics (SQMD),
based on a multi-species theorem of DFT for open quantum systems.
SQMD allows to treat both the electronic and nuclear degrees of
freedom open to environments, and, in principle, it provides all
possible dynamical correlations in the system. In particular, SQMD
takes into account energy relaxation and dephasing of the electronic
subsystem, a feature lacking in any ``standard'' MD approach. This
opens up the possibility to study a wealth of new phenomena such as
local ionic and electronic heating in laser fields, relaxation
processes in photochemistry, etc.

%%%%%%%%%%%%%%%%%%%%%%%%%%%%%%%%%%%%%%%%%%%%%%%%%%%%%%%%%%%%%%%%%%
%%%%%%%%%%%               Acknowledgement               %%%%%%%%%%
%%%%%%%%%%%%%%%%%%%%%%%%%%%%%%%%%%%%%%%%%%%%%%%%%%%%%%%%%%%%%%%%%%
We would like to thank Y. Dubi, R. Hatcher and M. Krems for useful
discussions.
Financial support by Lockheed Martin is gratefully acknowledged.

\end{document}